# Improvement of piezocatalytic performance of $Na_{0.5}Bi_{0.5}TiO_3$ perovskite using K doping for efficient Rhodamine B degradation

Salma Ayadh [1,2*], Salma Touili [1,2], M'barek Amjoud [1], Daoud Mezzane [1,2], Mohamed Gouné [3], Jaafar Ghanbaja [4], Manal Benyoussef [2], Hana Uršič [5,6], Nejc Suban [5,6], Mustapha Raihane [1,7], Zdravko Kutnjak [5], Mimoun El Marssi [2]

**1** Laboratory of Innovative Materials, Energy and Sustainable Development (IMED-Lab), Cadi Ayyad University, Faculty of Sciences and Technology, BP 549, Marrakech, Morocco.

**2** Laboratory of Physics of Condensed Matter (LPMC), University of Picardie Jules Verne, Scientific Pole, 33 rue Saint-Leu, 80039 Amiens Cedex 1, France.

**3** ICMCB, University of Bordeaux, 87 Avenue du Dr Albert Schweitzer, Pessac 33600, France.

**4** Univ Lorraine, CNRS, IJL, F-54000 Nancy, France.

**5** Jožef Stefan Institute, Jamova Cesta 39, 1000 Ljubljana, Slovenia.

**6** Jožef Stefan International Postgraduate School, Jamova Cesta 39, 1000 Ljubljana, Slovenia.

**7** Applied Chemistry and Engineering Research Center of Excellence (ACER-CeO), Mohammed VI Polytechnic University, Hay Moulay Rachid, 43150 Ben Guerir, Morocco.

*Corresponding author:

*E-mail:* s.ayadh.ced@uca.ac.ma

*ORCID:* https://orcid.org/0009-0004-6093-1680

## Abstract

Piezocatalysis, based on the piezoelectric properties of catalysts, breaks down the barrier between mechanical energy and chemical energy. It describes the use of charges induced by piezoelectricity to assist typical chemical processes while harvesting various forms of mechanical green energy. The performance of piezocatalysis is predominantly governed by the piezoelectric properties of materials. The main aim of this work is to evaluate and analyze the potential of potassium-doped sodium bismuth titanate $Na_{0.5-x}K_xBi_{0.5}TiO_3$ abbreviated as $NK_xBT$ (x=0, 0.15, 0.20, and 0.25), as a piezocatalyst in the degradation of the organic dye Rhodamine B (RhB) under ultrasonic vibration. The synthesis of $NK_xBT$ nanopowders was conducted using the sol-gel autocombustion method. Coupled structural analysis reveals the presence of an intermediate Morphotropic Phase Boundary (MPB; where two phases coexist) in the optimal $NK_{15}BT$ composition. The piezocatalytic degradation results showed a total piezo-degradation in only 90 min and a rate constant 8 times higher than the undoped $NK_0BT$. The enhanced piezocatalytic activity results from a synergistic effect of MPB presence, reduced particle size, optimal bandgap and high lattice strain. The $NK_{15}BT$ sample also demonstrated good reusability and good mineralization.

**Graphical Abstract**

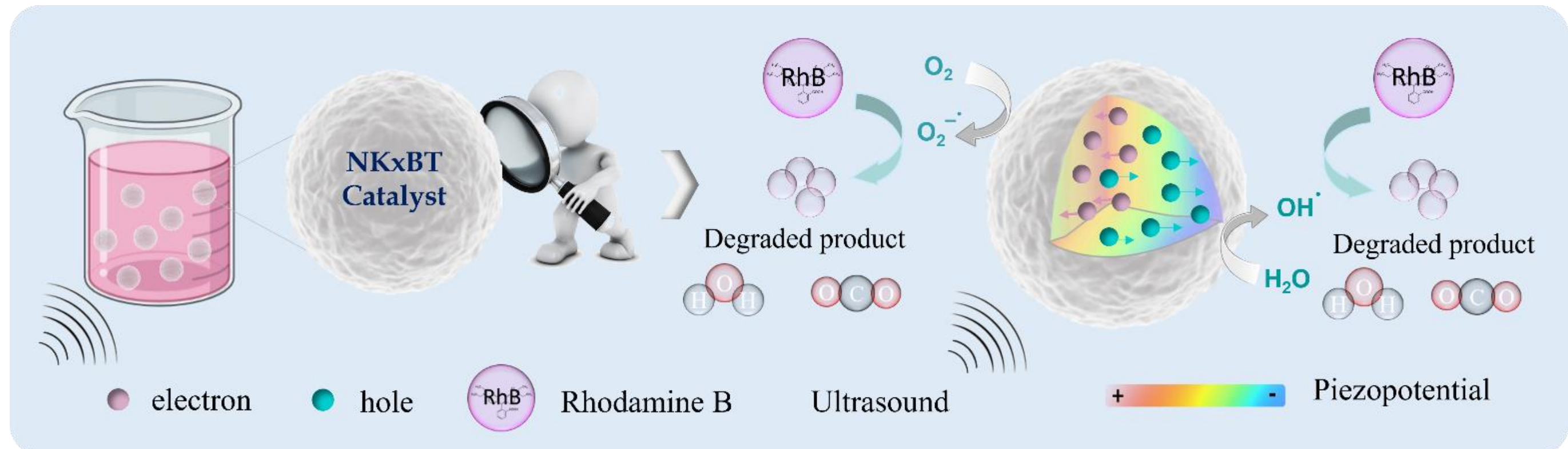

## Introduction

During the last decades, there've been a wide range of global threats that certainly require humanity's urgent attention. Climate change is well-known to be the most pressing challenge facing our planet, but it has often been perceived as a distant threat, with subtle impacts that can be easy to overlook in our daily lives. However, water is one of the primary ways through which we feel the effects of climate change manifesting in temperature extremes, floods, and droughts. While climate change decreases water quantity, population growth coupled with industrialization influence water quality by increasing emerging pollutants, pushing some regions to reach their freshwater resource limits. Among these pollutants, organic compounds, such as dyes, have widely used in numerous industries. However, despite regulatory norms, uncontrolled daily discharge of these hazardous dyes into natural water bodies, lead to severe water pollution. Over the past decade, many studies have been exploring different methods to treat water, specifically photocatalysis, which has been regarded as an ecofriendly approach, using light energy for pollutant degradation. Ongoing research has gradually introduced piezoelectric catalysis as a new approach in the field of wastewater treatment. Piezo-catalysis take the advantage of the piezoelectric effect and offer another source to be exploited to harvest green energy for dye degradation, which is mechanical energy. In recent years, many piezoelectric materials have been used for catalytic degradation, including perovskite titanates such as $(Ba,Sr)TiO_3$ [1], $Pb(Zr_{0.52}Ti_{0.48})O_3$ [2], $Ba_{1-x}Bi_xTi_{0.89}Sn_{0.11}O_3$ [3], alkali niobates $(Li,Na,K)NbO_3$ [4][5], $BiFeO_3$ [6]. Sodium Bismuth Titanate $Na_{0.5}Bi_{0.5}TiO_3$ (NBT) can be highlighted in virtue of the wide range of perovskite oxides materials due to its excellent piezoelectric properties such as strong ferroelectricity, large remnant polarization (~38 μC/cm$^2$), and high piezoelectric coefficient (~58 pC/N) [7]. There has been extensive studies and numerous efforts to improve its piezoelectric properties and its catalytic activity, where doping presents a promising strategy to achieve superior performance [8]. Considering this aspect, and searching for further enhancement, potassium (K) is chosen as a dopant in this work. As a highly polarized element, potassium can enhance piezoelectricity, hence, piezo-catalytic degradation [9].

In the present work, we synthesized different composition of K doped NBT ($NK_xBT$ x=0, 15, 20, 25%) by sol–gel auto-combustion method and explored their piezo-catalytic performances under ultrasonic vibration.

## 1.Experimental

### 1.1 Synthesis Method

Sodium Nitrate ($NaNO_3$, purity =99%), Bismuth Nitrate (Bi $(NO_3)_3 \cdot 5H_2O$, 98%), Potassium Nitrate ($KNO_3$, 99%), and Titanium butoxide ($Ti(C_8H_{17}O)_4$, 97%) were used as starting materials.

Firstly, stoichiometric amounts of bismuth nitrate were dissolved in distilled water and stirred on a magnetic stirrer until completely dissolved. Then Sodium Nitrate and Potassium Nitrate were added to the solution. Separately Titanium butoxide was dissolved in distilled water. The two solutions were mixed with a magnetic stirrer until a homogenous solution was obtained. Citric acid ($C_6H_8O_7$) was added as a fuel agent for the reaction at a 1:1 molar ratio. Using a hot plate, the solution was heated at 200°C with constant stirring until the formation of a gel. Subsequently, the sample is heated until the gel is burned out and the sample is transformed into ash. Then the resulting ash was calcined at 800°C for 6h to obtain pure $NK_xBT$ powders. After this treatment, the sample was ground into fine powder. The detailed synthesis procedure is shown in Figure 1. The four $NK_xBT$ samples with x=0, 15, 20, and 25 were synthesized and abbreviated as $NK_0BT$, $NK_{15}BT$, $NK_{20}BT$, and $NK_{25}BT$.

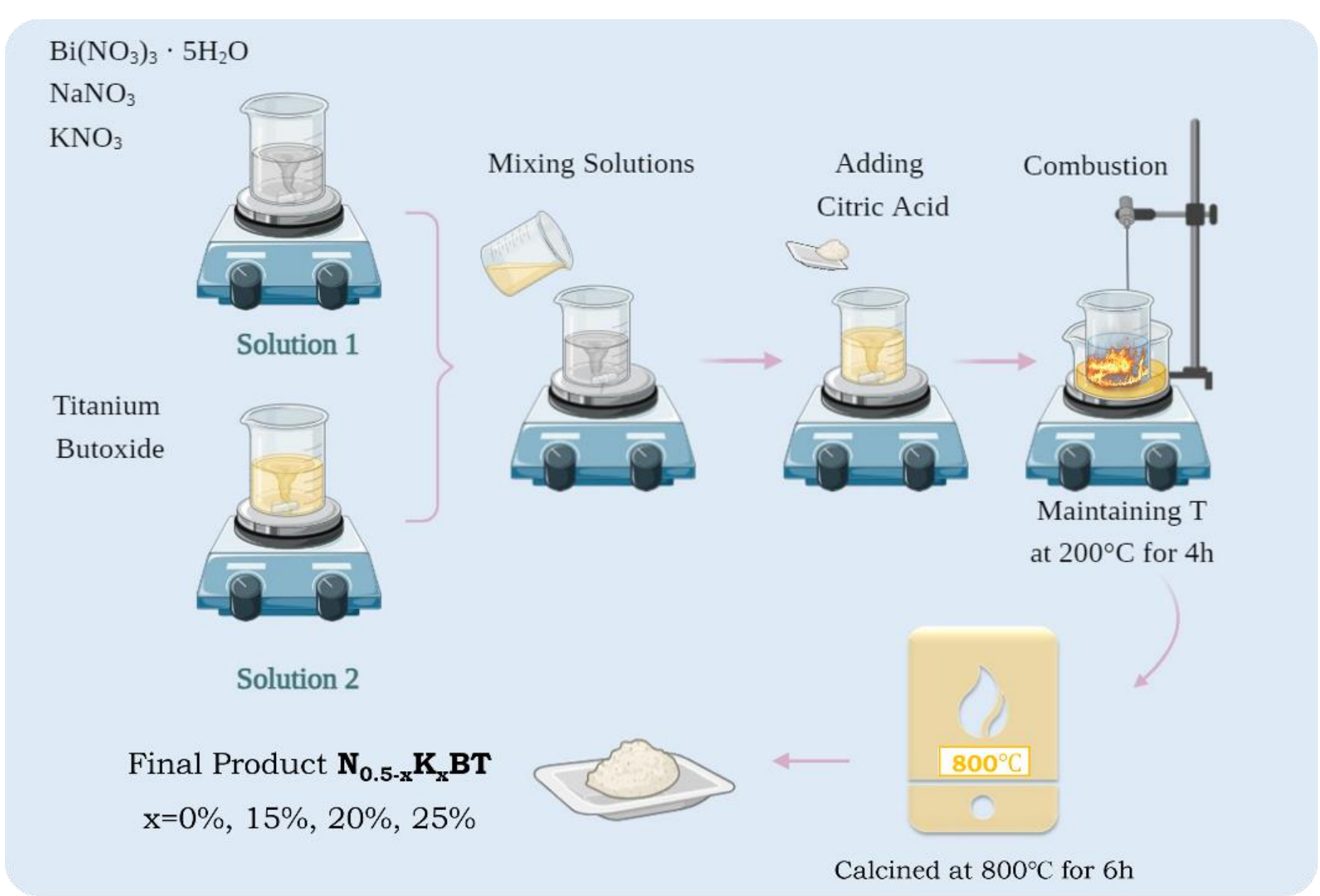

***Figure 1***. *The synthesis process of NKxBT powders by the sol-gel auto-combustion method.*

### 1.2 Instrumentation

The prepared powder's X-ray diffraction (XRD) data were recorded by"Rigaku" XRD-diffractometer with a Cu-Kα radiation source(λ=1.54Å). Transmission electron microscopy (TEM) investigations were carried out using a JEM- ARM 200F Cold FEG TEM/STEM operating at 200 kV and equipped with a spherical aberration (Cs) probe and image correctors (point resolution 0.12 nm in TEM mode and 0.078 nm in STEM mode). The Raman spectra were recorded using a Renishaw Raman spectrometer equipped with a CCD detector. A ZetaSizer-Malvern instrument was used for zeta potential measurements at 25°C, 2 mg of the powders of all samples were dispersed in 20 mL of distilled water. Visible spectra for piezo-degradation analysis were measured using a Shimadzu UV 2600 spectrophotometer for wavelengths ranging from 400 to 650 nm. The piezocatalytic performances of $NK_xBT$ samples were evaluated by the degradation of RhB dye (major absorption peak at 554nm). An ultrasonic with a frequency of 37 KHz and with a power of 300W was used to apply local mechanical strain to the samples. The piezo-catalytic tests were evaluated using10 mg of the sample dispersed in 10 ml of 5 mg/L RhB aqueous solution. Prior to ultrasonic vibration, the mixture solution was magnetically stirred in the dark for 120 min aimed to achieve adsorption-desorption equilibrium between RhB and sample. The RhB concentration after equilibration was denoted as the initial absorbance (A0). Subsequently, the mixture solution was exposed to ultrasonic vibration. In the degradation process, after every 30 minutes, the above mixture was collected and then centrifuged. The residual absorbance A of RhB at the degradation time was measured using a UV-vis spectrophotometer and plotted as a function of degradation time. The percentage of residual absorbance of RhB was indicated as A/A0. The degradation efficiency of the photocatalyst is defined in the following equation:

$$\eta = \frac{A}{A0} \times 100\ (Eq. 1)$$

## 2. Structural and Morphological characterization

### 2.1 X-ray diffraction

The X-ray diffraction (XRD) measurements were conducted to investigate the crystal structures of the prepared samples. It can be seen from Figure 2.a that the $K^+$ doped NBT crystal displays typical $ABO_3$ perovskite diffraction peaks for all compositions. However, a small peak around the 2θ range of 30° (marked by an asterisk) of secondary phase is found in the pure NBT and $NK_{15}BT$ diffractogram which is difficult to identify [10]. Based on the peaks position of the secondary phase, it can be attributed to $Bi_4Ti_3O_{12}$, $Na_2Ti_3O_7$ or $Na_{0.5}Bi_{4.5}Ti_3O_{15}$.

A deep examination of the XRD patterns in Figure 2.b reveals no obvious peak splitting existing in (110), (111) and (200), indicating that a predominantly pseudo-cubic phase was present for all compositions [11]. Closer inspection of (200) peak shows a very low intensity shoulder for NBT and $NK_{15}BT$ indicating that another phase may coexists with the majority of the pseudo-cubic phase. While, the pseudo-cubic phase is enhanced for composition with higher $K^+$ content ($NK_{20}BT$ and $NK_{25}BT$). This can be explained by the fact that $K^+$ is a larger and more polarizable ion compared to $Na^+$[12]. Substituting larger, more polarizable A-site ions ($K^+$) creates oversized cube-octahedral cages. This allows smaller, polarizable ions like $Bi^{3+}$ to shift off-center within the cages, generating local electric dipoles that stabilize the pseudo-cubic structure[9]. All the samples' structure can be better described in terms of pseudo-cubic symmetry, but, since larger views of (110), (111) and (200) reveal peak broadness, the possibility of a co-existence of other phases cannot be completely ruled out.

Moreover, the expanded XRD patterns reveal that the diffraction peaks tend to shift to the lower degrees with the K content increase, which suggests that the unit cell volume enlarged. This confirms the successful incorporation of $K^+$ into the NBT crystal lattice, because the enlargement of unit cell volume was caused by replacement of $Na^+$ (r =0.95 Å) by an element with larger radius $K^+$ (r = 1.33 Å) [13]. Table 1 shows that the introduction of K dopant in NBT results in an increase of lattice parameters a and in the unit cell volume.

The successful substitution of $Na^+$ ion by $K^+$ in the NBT lattice was also evident in a change in the microstrain and crystallite size. The average crystallite size was estimated by using both Debye–Scherer's formula (Eq. 2) and Williamson-Hall (W-H) method from the five most prominent diffraction peaks [100], [110], [111], [200], and [211]:

$$D = \frac{k\lambda}{\beta cos\theta} \ (Eq.2)$$

Where Scherrer constant k = 0.94 (for spherical crystallites with cubic symmetry), $\beta$: full width at half maximum (FWHM) of the most intense peak in radian, $\theta$: Bragg's diffraction angle, λ: wavelength of X-ray used (1.54178Å).

The microstrain has been calculated using the following formula (Eq.3):

$$\varepsilon = \frac{\beta_{hkl}}{4tan\theta} \ (Eq.3)$$

Where $\varepsilon$ is the strain factor.

In W-H method (Eq.4), by plotting $\beta_{hkl}$ cosθ against 4 $\beta_{hkl}$ sinθ for all the peaks and fitting the best line (Figure 3), both crystallite size and strain can be estimated, giving that the intercept and slope the fitting line are kλ d and ε, respectively [14].

$$\beta cos\theta = \varepsilon \ (4 \ sin\theta) + \frac{k\lambda}{D}) \ (Eq.4)$$

The dislocation density (δ), which is defined as the length of dislocation lines per unit volume (lines/$m^2$), is a measure of the number of defects in the grown nanocrystalline powder. It was calculated from the average values of the crystallite size (D) by the following equation (Eq.5) [15]:

$$\delta = \frac{1}{D^2} \ (Eq.5)$$

Where D is the crystallite size from the Debye–Scherrer equation.

The variation of crystallite size and microstrain as a function of the $NK_xBT$ samples is tabulated in Table 1 from which the inverse relation between microstrain and the size might be evidence.

It was found that the crystallite size estimated by both methods follow the same trend (Figure 4). It decreases from D:49 nm, W-H: 73.2 nm to a minimum value of D:20 nm, W-H: 29.2 nm for the $NK_{20}BT$. However, the opposite tendency is observed for higher K doping levels ($NK_{25}BT$). The reduction in crystallite size is mainly due to the occurrence of distortion in the NBT lattice due to the foreign atoms which creates more defects and reduces the nucleation and subsequent growth of nanoparticles caused by the drag force exerted by the dopant [15], [16]. By reducing the obstacles (dislocations) to grain boundary movement, the crystallite size can be larger. It is also worth noting to mention that the crystallite size determined using the W-H plot approach is larger than the value of crystallite size obtained using the Debye-Scherrer equation. This difference may be due to the fact that Scherrer method attributes the peak width only to the size of the crystallites. While the width of the peaks depends not only on the size of the crystallites but also on the strain of the lattice which is accounted for in W-H method. Therefore, W-H method can give a better estimate of the crystallite size [14].

Also due to this difference in ionic radius, the incorporation of potassium in the matrix during the chemical crystal growth process may lead to the formation of more structural disorders leading to internal stress in the host lattice that is maybe be the reason for the increase in the lattice strain for the two samples [17].

Then, the decrease of lattice strain at a higher doping concentration ($NK_{25}BT$) can also be attributed to the reduction of dislocation density [18]. Furthermore, the decrease of lattice strain could be ascribed to relaxation of the local strains when the introduced K atoms occupy the interstitial sites in the crystal lattice [19], even if the unit cell parameter "a" continues to increase. The reason for this unique behavior is believed to be a result of highly polarized K ions and weak K-O bonds [20]. At high doping concentrations, the effect of weakened bonding across multiple sites makes it easier for oxygen ions to detach from their lattice sites by lowering the energy required for oxygen ions to move within the lattice which led to significant vacancy formation. The presence of oxygen vacancies can redistribute stresses within the material, relax the lattice strain, and cause the surrounding atoms to move and accommodate the vacancy as well as the big size of the dopant ion, which can increase the lattice parameter [21].

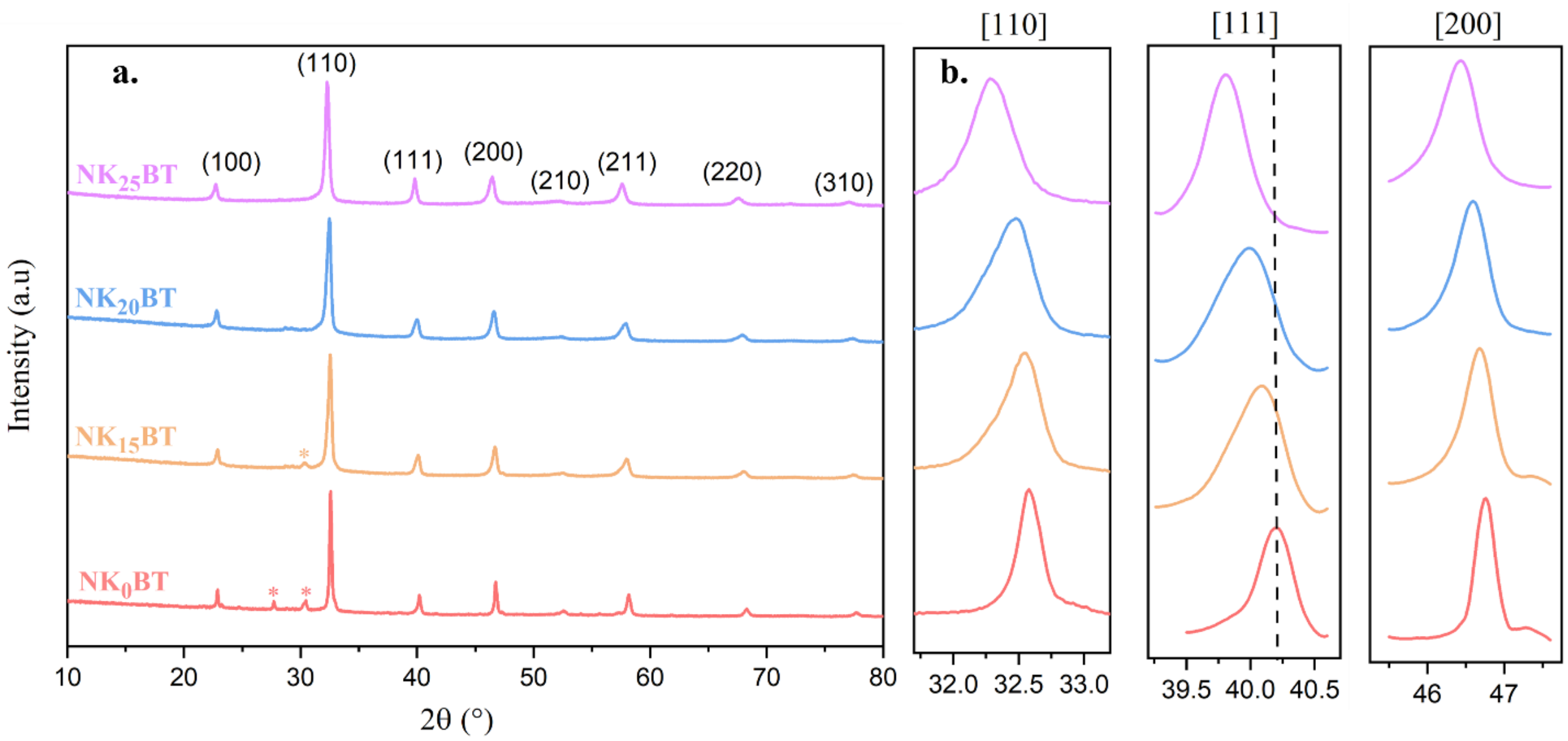

***Figure 2. a.** XRD pattern of NKxBT powders **b.** larger views of (110), (111) and (200) planes.*

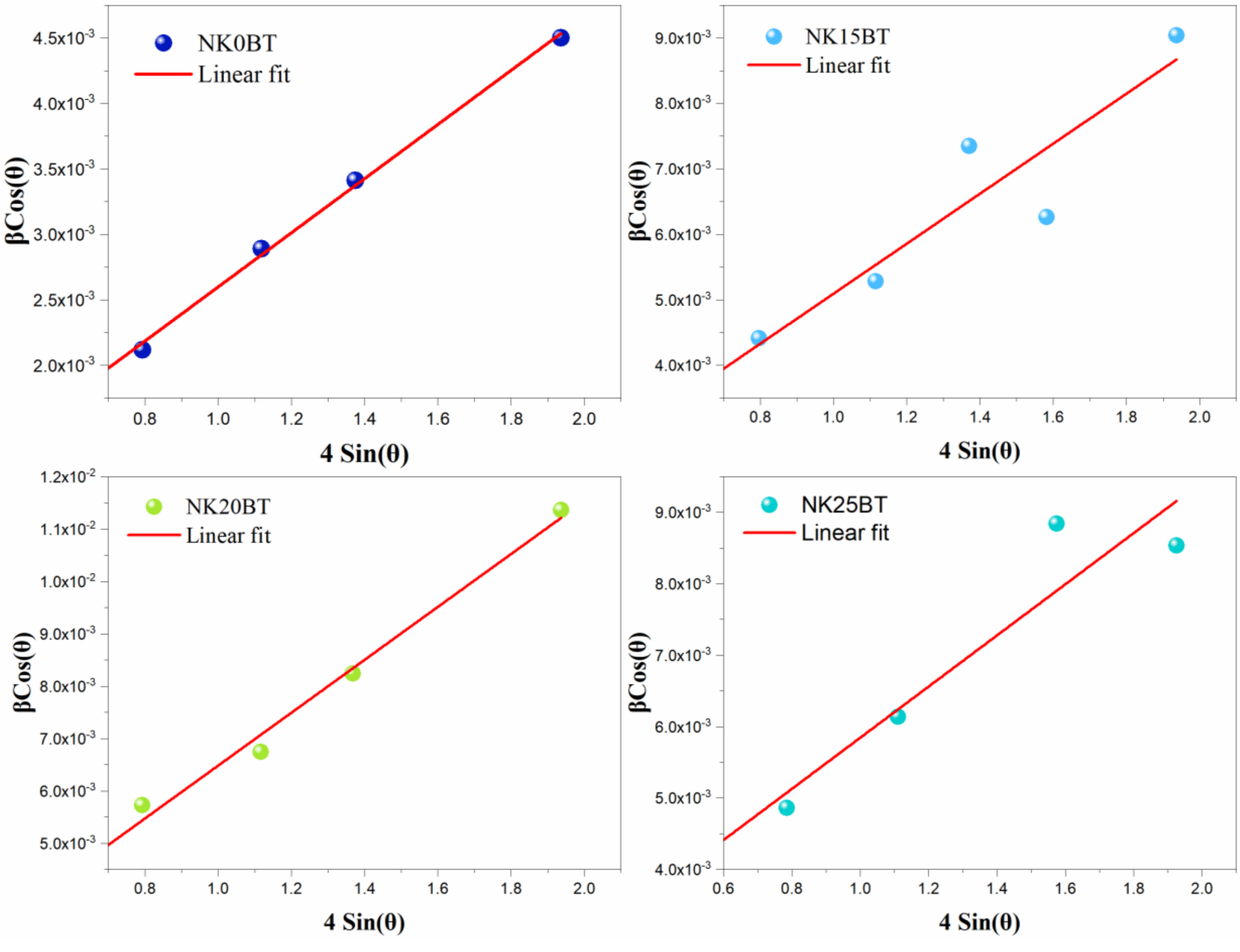

***Figure 3.*** *Williamson-Hall (W-H) plots using XRD data for $NK_xBT$ samples.*

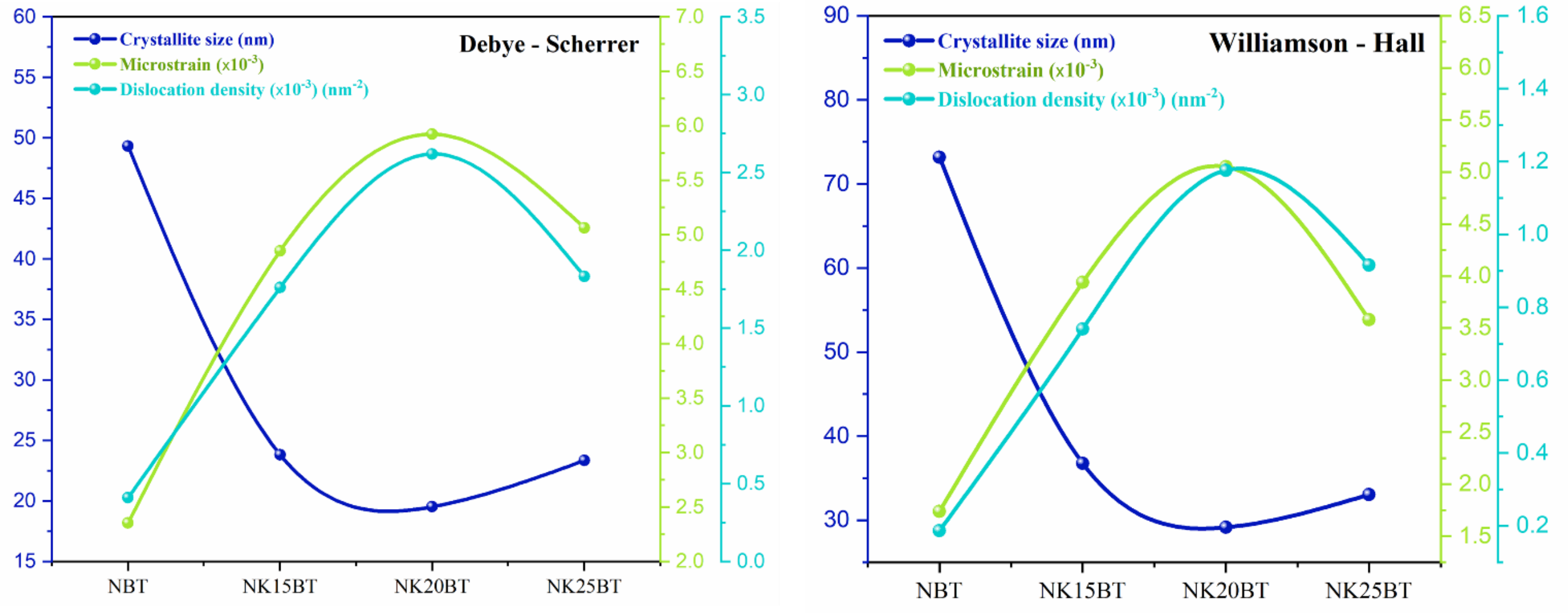

***Figure 4.*** *Variation of Crystallite size, dislocation density, and microstrain of NKxBT samples using Debye-Scherrer method (left) and William-Hall method (right).*

***Table 1.*** *Lattice parameters, cell volume of $NK_xBT$ samples.*

| $NK_xBT$ | a (Å) | V (Å³) | Crystallite size (nm) | | Microstrain (x $10^{-3}$) | | Dislocation density $\delta$ ($nm^{-2}$) (x$10^{-3}$) | |
|---|---|---|---|---|---|---|---|---|
| | | | D | W-H | D | W-H | D | W-H |
| $NK_0BT$ | 3.88 | 58.4 | 49.3 | 73.2 | 2.3 | 1.7 | 0.4 | 0.2 |
| $NK_{15}BT$ | 3.89 | 58.9 | 23.8 | 36.7 | 4.8 | 3.8 | 1.7 | 0.7 |
| $NK_{20}BT$ | 3.90 | 59.5 | 19.5 | 29.2 | 5.9 | 5.0 | 2.6 | 1.2 |
| $NK_{25}BT$ | 3.91 | 60.1 | 23.3 | 33.0 | 5.1 | 3.6 | 1.8 | 0.9 |

## 2.2 TEM microscopy

As a long-range technique, XRD reveals only the average periodic structure of the material [22]. However, in the case of a material with local structural variations or multiple phases, it is mandatory to complement this technique with electron diffraction and microscopy. Thus, in this work, High-resolution Transmission Electron Microscopy (HRTEM) is used for constructing more details of crystal structure, phase identification, morphology and element distributions via selected area electron diffraction (SAED) and High Angle Annular Dark Field Scanning Transmission Electron Microscopy (HAADF-STEM).

As can be seen from HRTEM images (Figure 5.a), all $NK_xBT$ samples tended to have cuboid-like morphology with some irregularly shaped domains composed of aggregations. The variation average size of $NK_xBT$ grains, which was determined by counting the number of grains across the diagonal of approximately one hundred grains using Image-J software, is shown in Figure 5.b. As can be observed from the figure, the average size decreases from 331 $nm$ ,with K-doping, reaches a minimum value of 134 $nm$ ($NK_{20}BT$). This can be attributed to the incorporation of larger K-ions into the NBT lattice, which can act as pinning site that prevent grain boundaries from moving and inhibit mass transit, thereby resulting in grain size refinement [23]. However, the trend of increase in particle size as evidenced from XRD for higher concentrations of $K^+$ is followed in TEM analysis. Further addition of K ions resulted in an increase of the average grain size and reached 207 $nm$ at the $NK_{25}BT$. As aforementioned, for higher K doping levels a strain relaxation may occurs in the crystal lattice. This relaxation of strain may facilitate grain boundary movement, allowing for increased grain growth [24].

The grain sizes observed in TEM images demonstrate that the particles are of submicron size, which is typically larger than the crystallite sizes estimated using Debye-Scherrer and W-H equations from XRD data. This difference arises mainly from the fact that crystallite size represents the average size of individual crystalline domains, whereas grain size refers to the average size of clusters of these crystalline domains. Generally, grains are composed of multiple crystallites. Also, TEM provides direct imaging of individual particles, capturing their actual dimensions, which may include agglomeration of these small crystallites. Moreover, the presence of defects within the material can affect XRD peak broadening, which may not accurately reflect the true dimensions of particles as seen in TEM. This can lead to an underestimation of crystallite size when using the Debye-Scherrer and W-H methods which focus on coherent scattering regions rather than overall particle morphology.

The SAED analyses performed locally along different zone axis are illustrated in Figure 6.b. The bright and well-defined diffraction spots in the SAED pattern confirm the high crystallinity of $NK_xBT$ samples[25]. In addition, from the SAED images obtained along the [001] and [101] zone axis orientations and the measured d-spacings, all the samples' patterns are indexed according the pseudo-cubic structure, supporting the XRD results. Furthermore, images from HAADF-STEM and their corresponding elemental mappings reveal a homogeneous distribution of all elements in the $NK_xBT$ samples (Figure 6.c).

To investigate the second phase, SAED pattern along the zone axis [130] of $NK_0BT$ has been shown in Figure 7. It can be seen that the d-spacings can be matched to planes in orthorhombic structure, with $Cmc2_1$ symmetry, which suggest its coexistence with the observed pseudo-cubic phase. In addition, the corresponding EDS elemental distribution (Figure 7.b), reveals that the elements of Bi and Ti gathered in the second phase, but Na element is relatively rare. Thus, the second phase can be attributed to a sodium-depleted compound $Na_{0.5}Bi_{4.5}Ti_3O_{15}$.

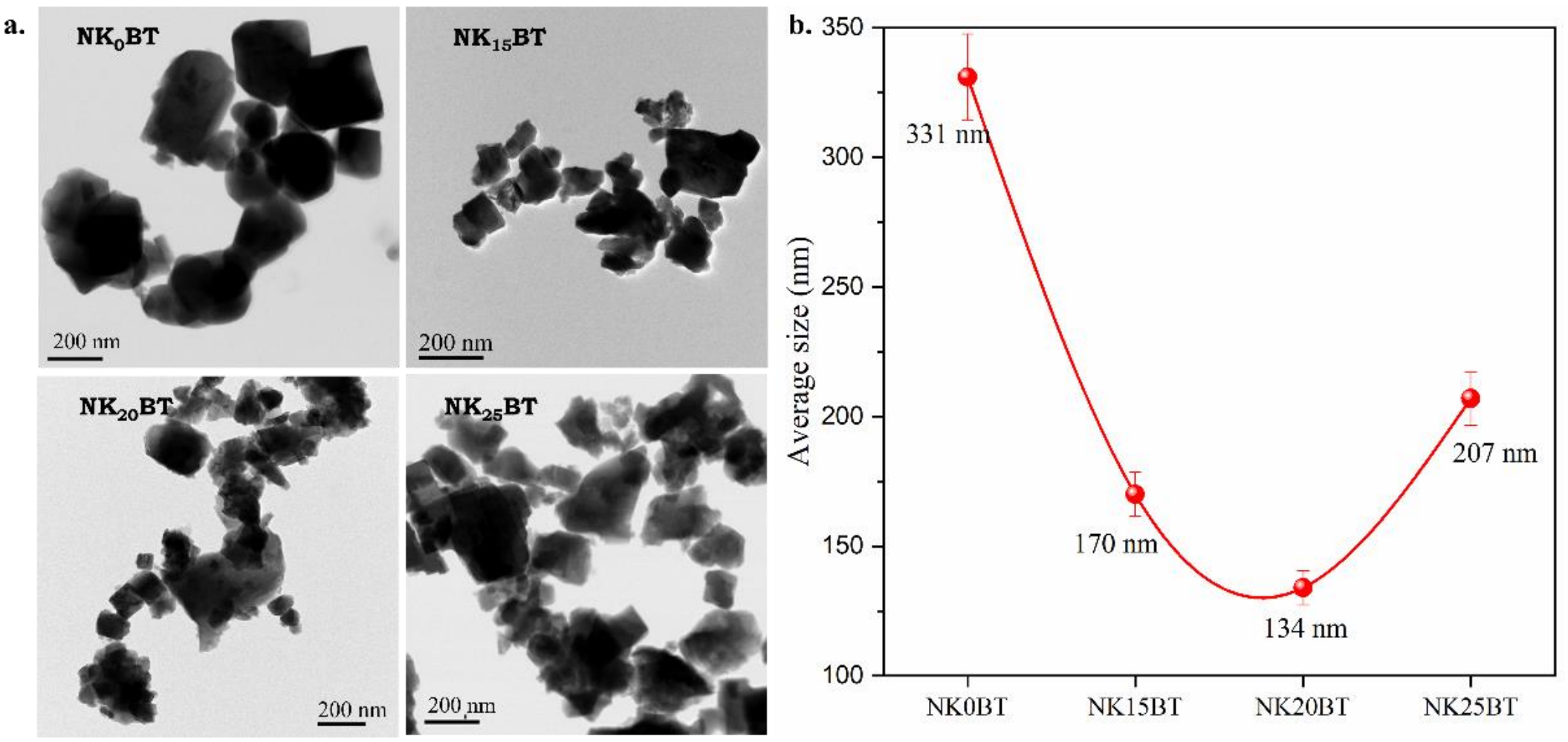

***Figure 5.a.*** *TEM images of $NK_xBT$ powders,* ***b.*** *Variation of average grain size of $NK_xBT$ powders.*

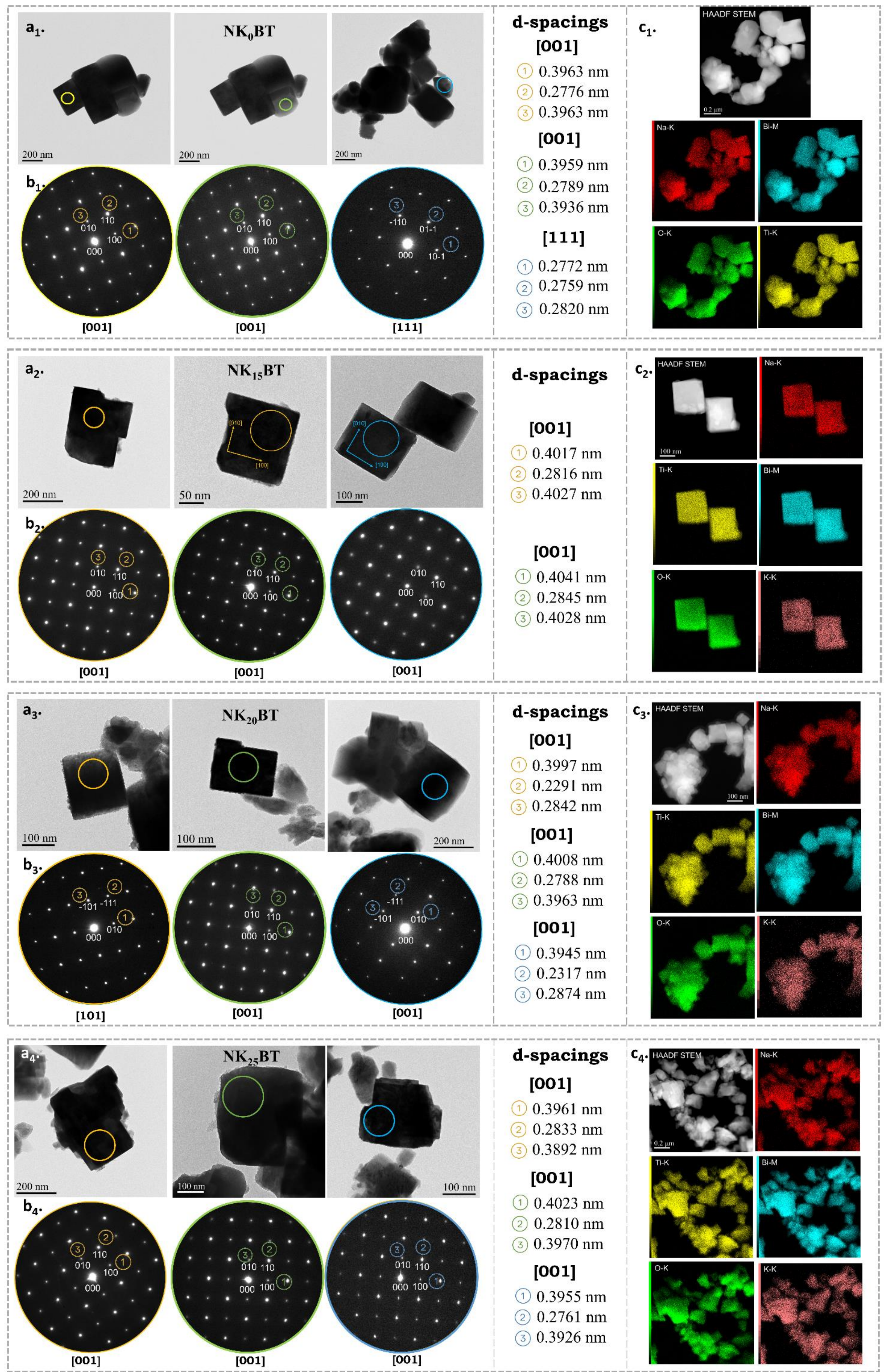

***Figure 6. a.*** *TEM images,* ***b.*** *SAED paterns,* ***c.*** *AADF STEM images and TEM mapping of NKxBT powders.*

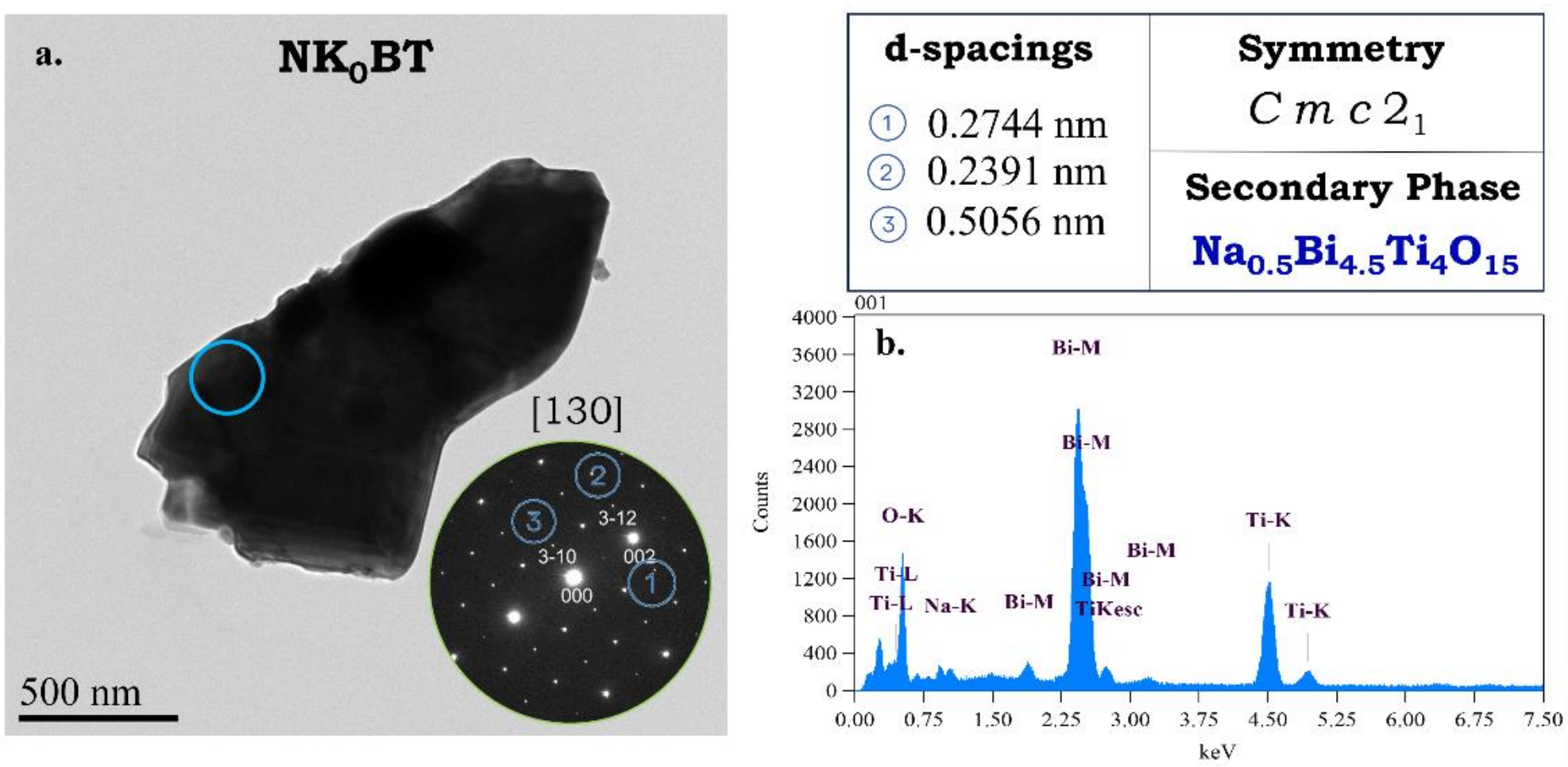

***Figure 7. a**. TEM image, **b.** EDS of the secondary phase in $NK_0BT$.*

**2.3 Raman spectroscopy**

As a long-range technique, XRD reveals only the average periodic structure of the material [37]. However, in the case of a material with local structural variations or multiple phases, it is mandatory to complement this technique with another one that can detect local symmetry. Raman spectroscopy has been investigated as a sensitive technique, even to very small phase fractions [38]. Raman spectra of $NK_xBT$ samples, in the range of 15-1100 $cm^{-1}$ at room temperature, were represented in Figure 3. As can be seen, the spectral bands are relatively broad, which can be primarily due to the simultaneous presence of $Na^+$, $K^+$ and $Bi^{3+}$ ions causing overlap of some Raman bands [39].

Four regions can be detected (Figure 3.a), which is in agreement with previous literature studies on NBT-based materials [40]. The low wavenumber region (15–200 $cm^{-1}$) corresponds to the vibrations of A-site cations. The midwavenumber range (200–400 $cm^{-1}$) is attributed to vibrations of the Ti–O bond. The high wavenumber region (450–650 $cm^{-1}$) is associated with the vibrations of the $TiO_6$ octahedra. Then the modes at 765 and 865 $cm^{-1}$, were mainly related to the longitudinal optical overlapped peaks $A_1$ and E [41]. Based on some literature studies, the two latest modes can be attributed to a change in oxygen vacancy density [42]. The detected bands in each region are characteristic of the rhombohedral NBT phase [43].

For the first region, at 25 $cm^{-1}$ and 134 $cm^{-1}$, a slight shift to lower frequencies is observed with increasing potassium doping ratio (Figure 3.b). The band position shift can be mainly caused by the increasing average mass on the A site, as K ions have a larger mass compared with that of Na ions [44].

The second and third regions, ascribed to the Ti-O bond and $TiO_6$ octahedra, respectively, show that by introducing K ions in the NBT phase, the modes at 265 and 536 cm$^{-1}$ start splitting into two bands at 320 and 625 cm$^{-1}$, respectively. The peak splitting of the $NK_xBT$ samples becomes more evident with a further increase in x. It is well known that, at room temperature, NBT crystallizes with a rhombohedral symmetry (R3c), whereas KBT crystallizes with a tetragonal symmetry (P4mm)[45]. Hence, these Raman modes suggest a change in symmetry to the tetragonal phase, which is in accordance with the XRD results [45]. Consequently, it can be suggested that the MPB region occurs in the $NK_{15}BT$ sample.

To give clear evidence for the formation of the morphotropic phase boundary at $NK_{15}BT$, Raman spectra of $NK_{15}BT$ and $NK_{25}BT$ compositions at low temperatures from -190 to 25℃ and at elevated temperatures from 25 to 450℃ were recorded (Figure S4). The spectra were corrected for Bose-Einstein statistics [n($\omega$)+1] to account for the intensity change caused by temperature variation [10].

Both spectra exhibit variations in peak intensities as the temperature decreases. However, the $NK_{25}BT$ Raman spectra show sharper and more distinct peak splitting (at 265 and 536 cm$^{-1}$), suggesting a well-defined tetragonal structure. While the peak intensities of $NK_{15}BT$ evolve smoothly with decreasing temperature, without clear transitions, retaining the characteristic bands of both structures, rhombohedral and tetragonal [12].

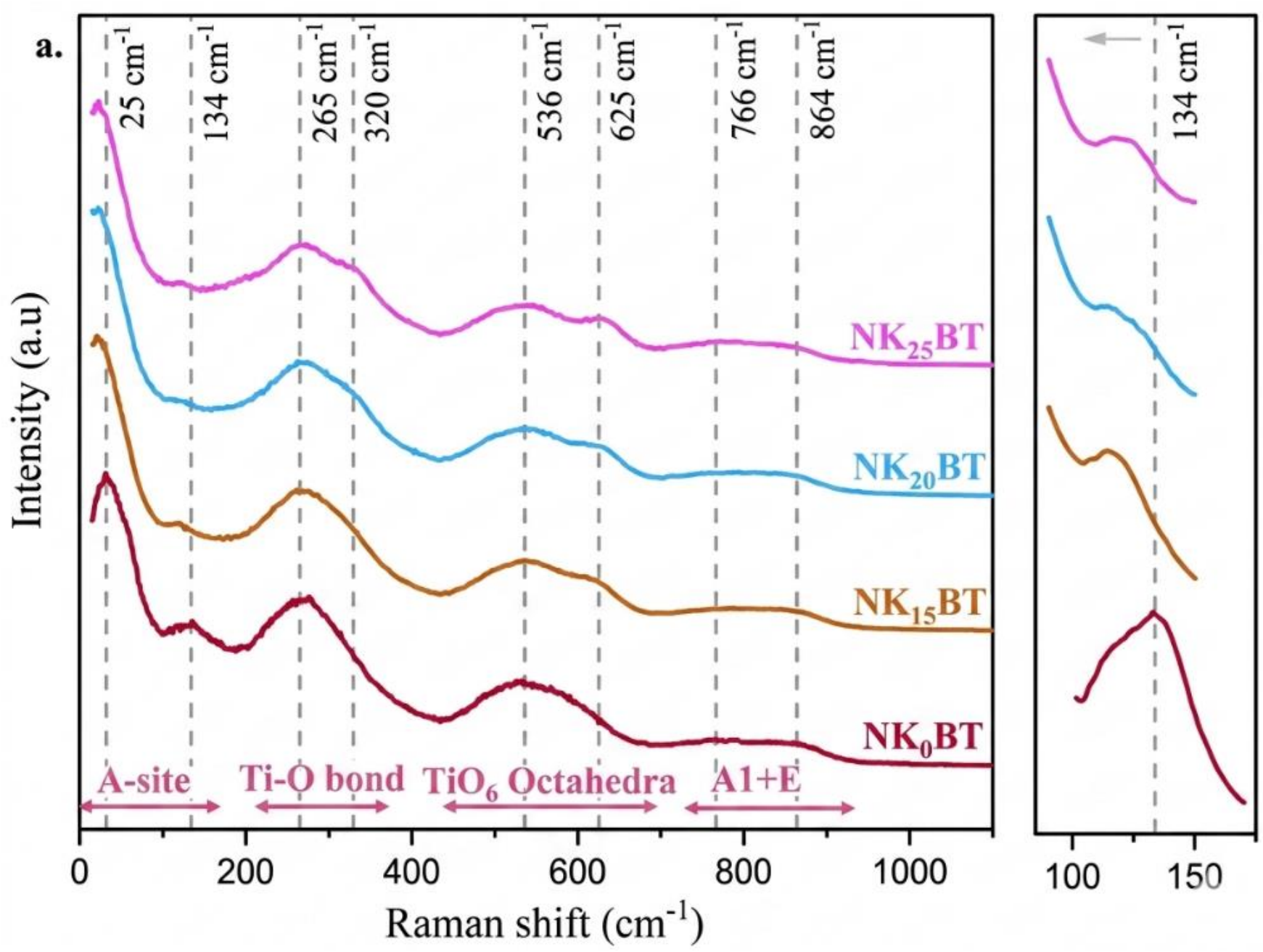

***Figure 8. a.*** *Raman spectra of NKxBT powders at ambient temperature,* ***b.*** *Larger view of the peak at 134 cm$^{-1}$.*

**3. Zeta-potential results**

As the primary goal in this work is to investigate the catalytic activity of our materials, it is crucial to examine their surface charge behavior at different pH values. Understanding how the surface charge varies with pH will allow us to determine the optimal conditions under which the degradation tests should be conducted. By selecting a pH where the interaction between the photocatalyst and the dye is not dominated by adsorption, but rather facilitates true photo-piezocatalytic degradation, we can ensure the accuracy of our results.

The dependence of surface charge of the $NK_xBT$ catalysts on pH determined using zeta potential measurements is shown in Figure 9. The point of zero charge ($pH_{pzc}$) is found between a pH value of 2.3 and 3.8. At a $pH < pH_{pzc}$, the surface of samples is positively charged, while at $pH > pH_{pzc}$, the surface carries a negative charge. It is well known that the carboxyl group of RhB has a pKa of 3.7, which leads the dye to adopt a cationic form (positively charged diethylamino group ($C{=}N^+$) when pH is below 3.7 and a zwitterionic form (due to the dissociation of the carboxylic group $COOH \rightarrow COO^-$) [35] when pH exceeds 3.7. Hence, in the context of the photo-piezocatalytic degradation tests, the experimental conditions involved maintaining a pH value of 6. At this pH, RhB is neutral, and the $NK_xBT$ samples demonstrated negative surface charge within the range of -20.66 to -26.88 mV. Under this condition, we can confirm that any observed decrease in dye concentration is due to actual degradation and not merely due to adsorption onto the material's surface.

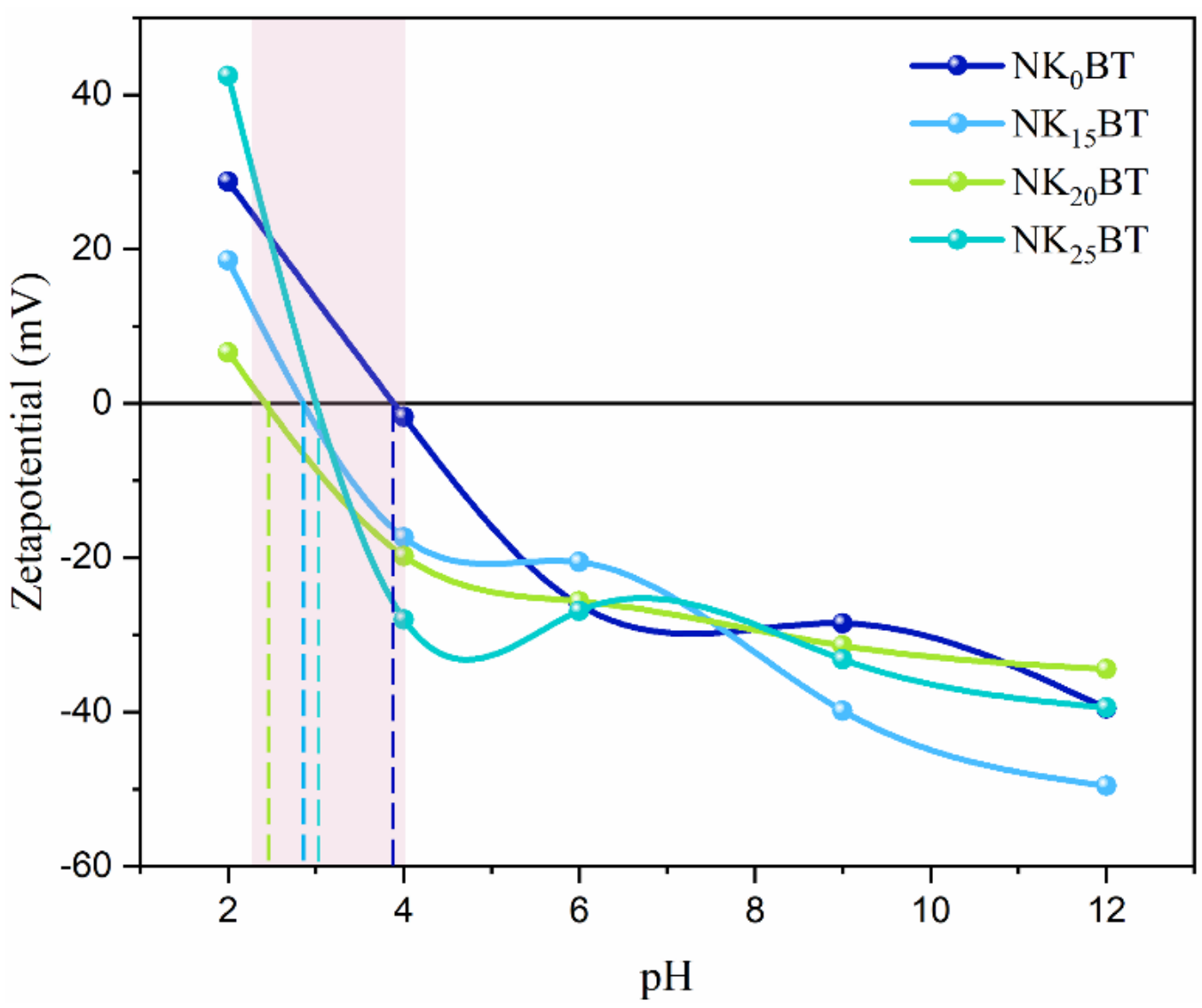

***Figure 9.*** *Variation of zetapotential of $NK_xBT$ powders at different pH values.*

**4. Bandgap**

While the band gap is primarily significant in photocatalysis, it still plays an indirect but important role in piezocatalysis as well. Materials with an optimal band gap will show better charge separation and mobility when subjected to mechanical stress.

To calculate the band gap energy at different K-doping levels, the following Tauc's relationship (Eq3) was used [4]:

$$(\alpha . h\nu)^{\frac{1}{\gamma}} = \mathrm{B}(h\nu - Eg) \qquad (Eq3)$$

Where $h$ is the Planck constant, ν is the photon's frequency, $E_g$ is the band gap energy, and B is a constant. The γ factor depends on the nature of the electron transition and is equal to 1/2 or 2 for the direct and indirect transition band gaps, respectively. Therefore, the energy gap ($E_g$) deduced from the x-intercept of the linear extrapolation of the curve $(\alpha.h\nu)^2$ as a function of photon energy radiation ($h\nu$) as shown in Figure S9. However, Figure 6.a shows the variation of the energy gap in $NK_xBT$ nanopowders, where the energy gap first decreases to a minimum of 2.75 eV for $NK_{15}BT$, then increases with further doping reaching 2.85 eV and 2.95 eV for $NK_{20}BT$ and $NK_{25}BT$, respectively. The introduction of dopant ions ($K^+$) can create new energy levels within the band structure; thus, the conduction band can be lowered, which explains the narrowing of the energy gap for $NK_{15}BT$ [52]. However, an increase in higher K-doping concentration results in a screening effect that hinders electronic transitions to the conduction band, leading to an increase in the optical band gap [53].

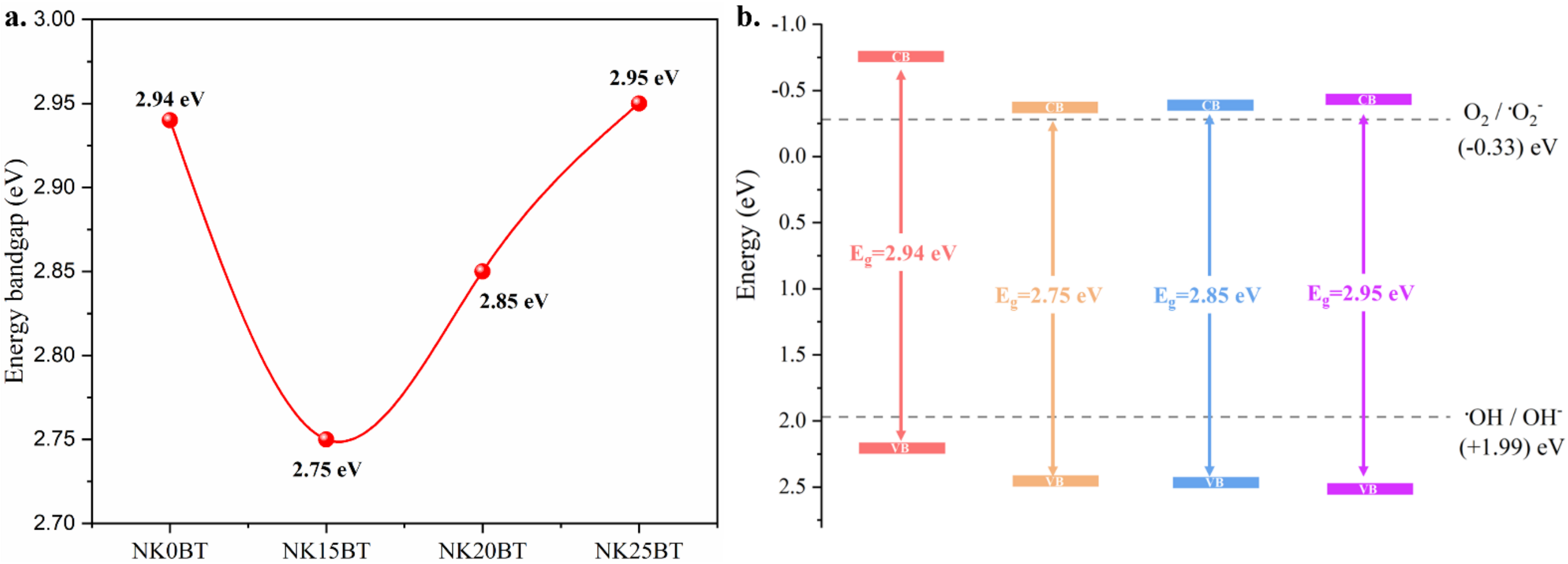

***Figure 10.a.*** *Variation of bandgap values of $NK_xBT$ powders* ***b****. Bandgap edges positions of $NK_xBT$ powders.*

It is well known that the observed change in the optical band gap for different compositions results from a shift in the positions of the conduction band (CB) and the valence band (VB). To illustrate this shift, the CB and VB were calculated using the following equations [4]:

$$E_{VB} = \chi - E_0 + 0.5Eg \quad (Eq4)$$

$$E_{CB} = E_{VB} - Eg \quad (Eq5)$$

Where $E_{VB}$ and $E_{CB}$ are the valence band and conduction band potentials, respectively, $E_0$ represents the energy of the free electrons versus hydrogen, which equals 4.5, while χ is the electronegativity of the semiconductor, which is determined using the following equation:

$$\chi = [\chi(A)^{a}\chi(B)^{b}\chi(C)^{c}\chi(D)^{d}\chi(E)^{e}]^{\frac{1}{a+b+c+d+e}} \qquad (Eq6)$$

Where the parameters *a*, *b*, *c*, *d*, and *e* are the number of atoms in each composition, and χ(*y*) is the electronegativity of each element of the compound, was derived from Pearson's research work [54]. Therefore, the electronegativity was calculated to be 5.223 eV, 5.552 eV, 5.543 eV, and 5.534 eV for $NK_0BT$, $NK_{15}BT$, $NK_{20}BT$, and $NK_{25}BT$, respectively. Based on these values, the calculated $E_{CB}$ and $E_{VB}$ edge potentials for $NK_0BT$, $NK_{15}BT$, $NK_{20}BT$, and $NK_{25}BT$ are as follows: (-0.747 eV, 2.193 eV), (-0.350 eV, 2.457 eV), (-0.382 eV, 2.468 eV), and (-0.436 eV, 2.504 eV), respectively (Figure 6.b).

The positions of CB and VB can serve as a good indicator of the reduction and oxidation abilities of generated electrons and holes, respectively. For effective degradation, the CB must be more negative than $O_2/\bullet O_2^-$ (-0.33 eV) and the VB more positive than $H_2O/\bullet OH$ (+1.99 eV) [55]. Typically, an optimal piezocatalyst band gap must be narrow enough to allow the excitation of charges by mechanical energy and wide enough to generate ROS with sufficient redox power [56]. All $NK_xBT$ compositions meet these requirements, confirming their ability to generate both $\bullet O_2^-$ (via electrons) and •OH (via holes) for efficient pollutant degradation; however, $NK_{15}BT$ nanopowder exhibits the narrowest energy gap, which facilitates greater charge carrier excitation.

**5. Adsorption/Desorption study**

The absorption spectra of RhB solution for varying immersion times (0, 60, and 120 min) of $NK_xBT$ samples in dark conditions are shown in Figure 11. For all samples, the absorbance of rhodamine B decreases during the first 60 minutes of dark conditions. However, there is no obvious change in the absorbance during the 60 to 120 min period under the same conditions. This indicates that the adsorption–desorption equilibrium could be reached after 60 min in the dark for all samples.

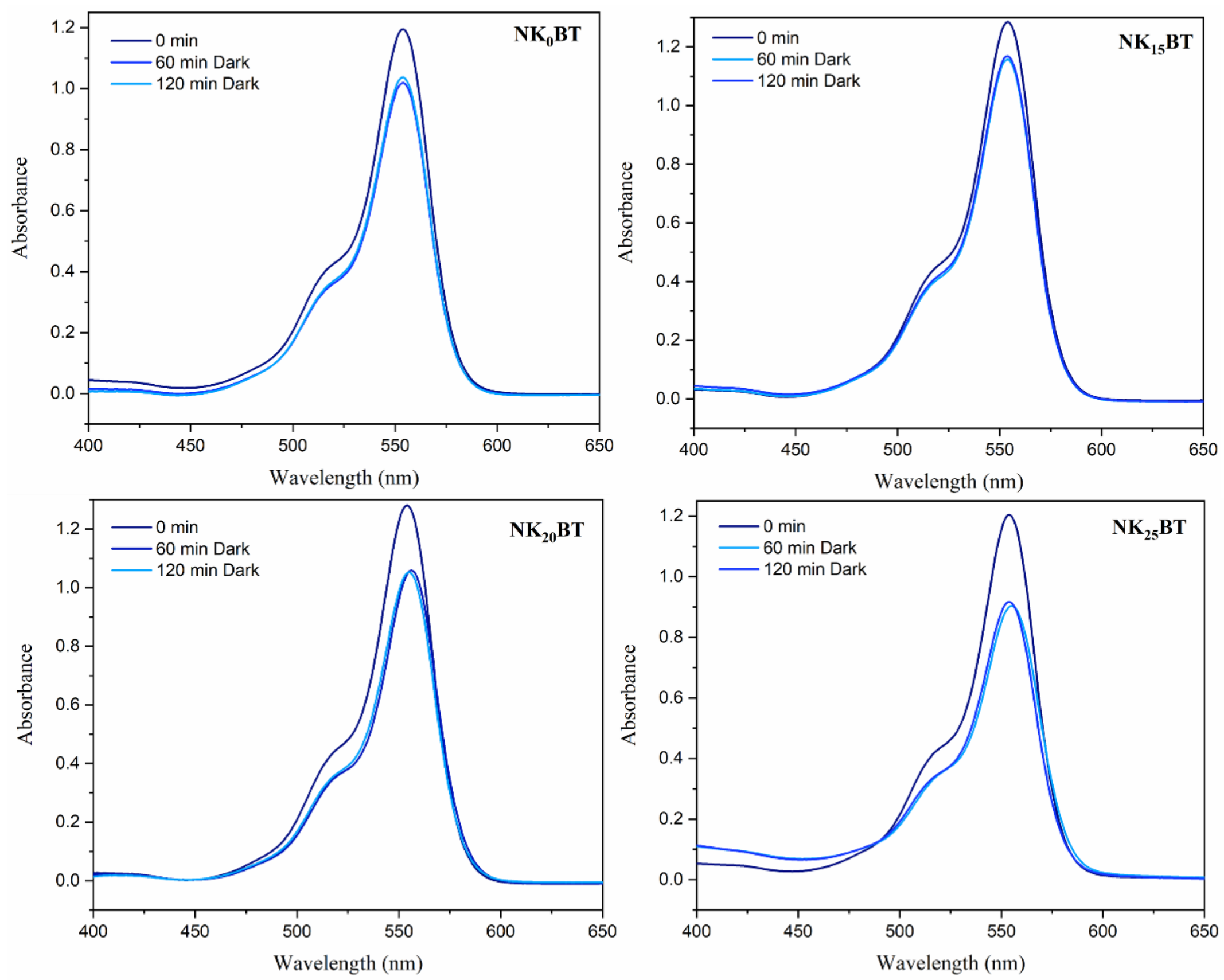

***Figure 11.*** *UV-Vis spectra of RhB solution under dark conditions.*

## 6. Piezo-catalytic activity of NK$_x$BT

The piezocatalytic capabilities of NK$_x$BT powders were studied under ultrasound to catalytically degradation of RhB. From Figure 12, it is evident that all the compositions showed some capability to degrade RhB; in particular the compositions NK$_{15}$BT and NK$_{20}$BT exhibit faster degradation as the exposure time was increased. After 90 min and 210 min of exposure, the absorption at 554 nm became almost zero, indicating that all RhB had been degraded by NK$_{15}$BT and NK$_{20}$BT piezo-catalysts.

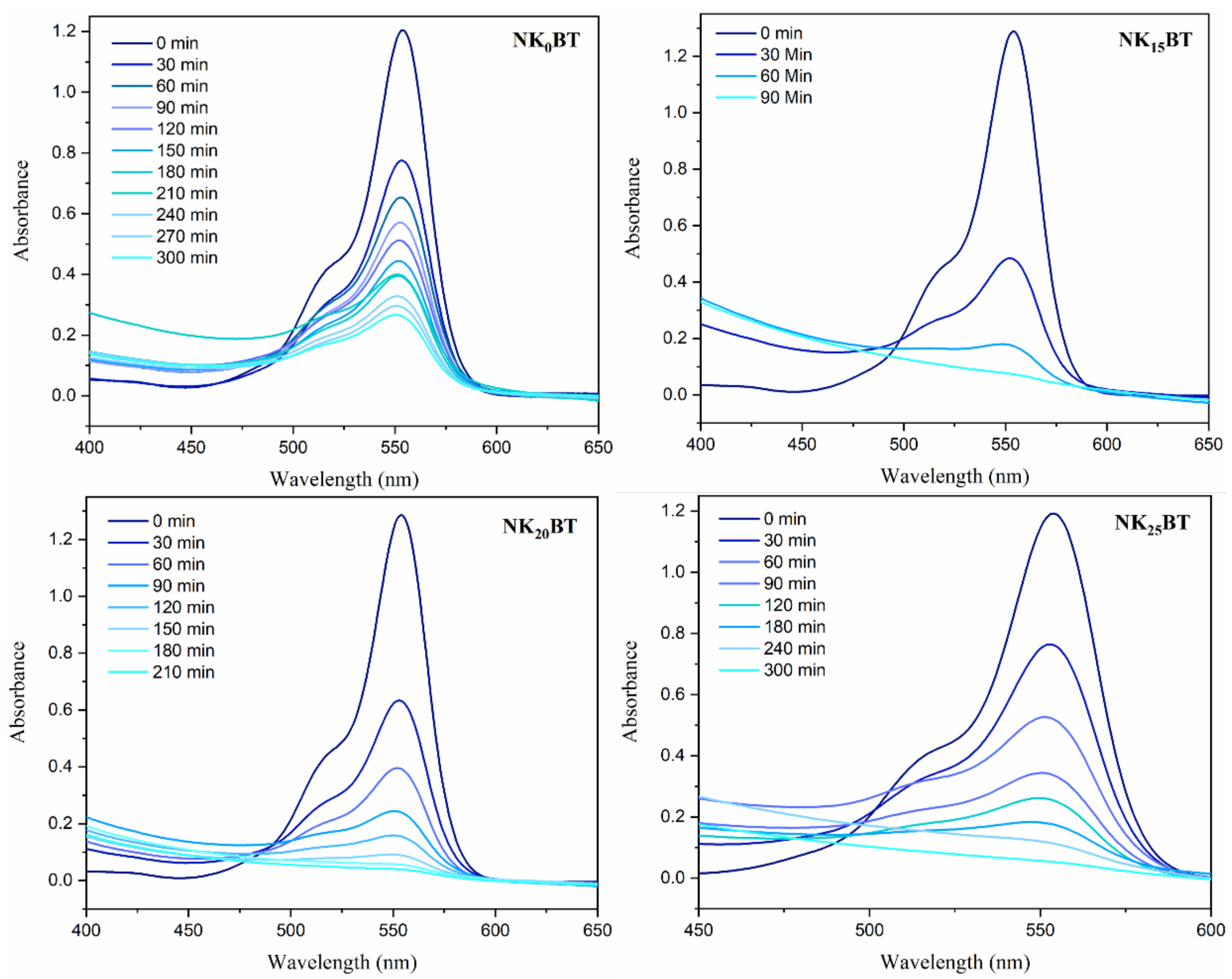

***Figure 12.*** *UV-Vis spectra of RhB solution under ultrasound vibration.*

## 7. Kinetic study

As shown in Figure 13.a, the blank test without a catalyst has been conducted and reveals that the self-degradation of RhB under ultrasound vibration occurs very slowly, and it is much less than the catalytic degradation of over a catalyst, which can be neglected. While, the increase in the efficiency of the piezo-catalytic degradation of RhB becomes noticeable after doping NBT with potassium. It was found that the degradation rate of RhB reaches 45.3 %, 99,8 %, 76.1 % and 65.8 %, for $NK_0BT$, $NK_{15}BT$, $NK_{20}BT$, and $NK_{25}BT$, respectively, while undergoing an ultrasound vibration for 90 min, demonstrating a notable improvement (Figure 13.b).

This improvement can be attributed to the generation of additional free charges on the surface when the catalysts are exposed to ultrasound waves [45]. Ultrasound also creates a polarized electric field inside the materials which generates a driving force [46]. As a result, the free charges accumulate continuously in the conduction band (CB) and valence band (VB) of piezo-catalysts, and causes shifts in the CB and VB. This shift improves the separation of charges (electrons and holes), preventing them from recombining, thus enhances their piezo-catalytic activity [45]. Additionally, their smaller particle size compared to $NK_0BT$ likely increases the specific surface area of the catalysts, allowing for enhanced interaction between the catalyst and RhB molecules, thereby boosting overall piezo-catalytic performance [47].

It was also reported, that the enhanced pseudo-cubic phase present advantages over catalytic activity. It has been suggested that this phase is favorable for improving the electron−hole separation [48], [49]. Moreover, the lattice strain of K doped NBT can also contribute to this improvement, giving that higher lattice strain significantly improve piezoelectricity, hence improve charge carriers' separation and piezo-catalytic activity [50].

As shown in Figure 14.b, $NK_{15}BT$ exhibits the highest piezo-degradation efficiency (99.8 %). This can mainly due to its small average grain size (0.17 $\mu m$) coupled with small band gap value (2.34 eV). Smaller particle size increases the specific surface area of the catalyst, and smaller band gaps facilitate easier electron excitation (even from mechanical energy), potentially allowing for better mobility which can lead to enhanced piezo-catalytic activity. Also, the simultaneous local presence of the different phases (rhombohedral-tetragonal-pseudo-cubic) in $NK_{15}BT$ as indicated by Raman spectroscopy, can lead to a mutation in the piezoelectric properties, which could enhance the piezo-degradation activity [45].

In order to determine the kinetic regularity for the piezo-degradation of RhB by $NK_xBT$ samples, the pseudo-first-order kinetic model (Eq.10), was used to fit the as-obtained data plots (Figure 13.c), which can be expressed as [51]:

$$-\ln\left(\frac{A}{A0}\right) = kt \quad (Eq.10)$$

The calculated rate constants are shown in Figure 13.d. Notably, the degradation rate over $NK_{15}BT$, $NK_{20}BT$ and $NK_{25}BT$ was about 8, 3 and 2 times of pure $NK_0BT$, respectively, and the maximum degradation rate constant is 0.036 $min^{-1}$, which shows a good degradation efficiency through comparing the different piezoelectric materials listed in Table 2.

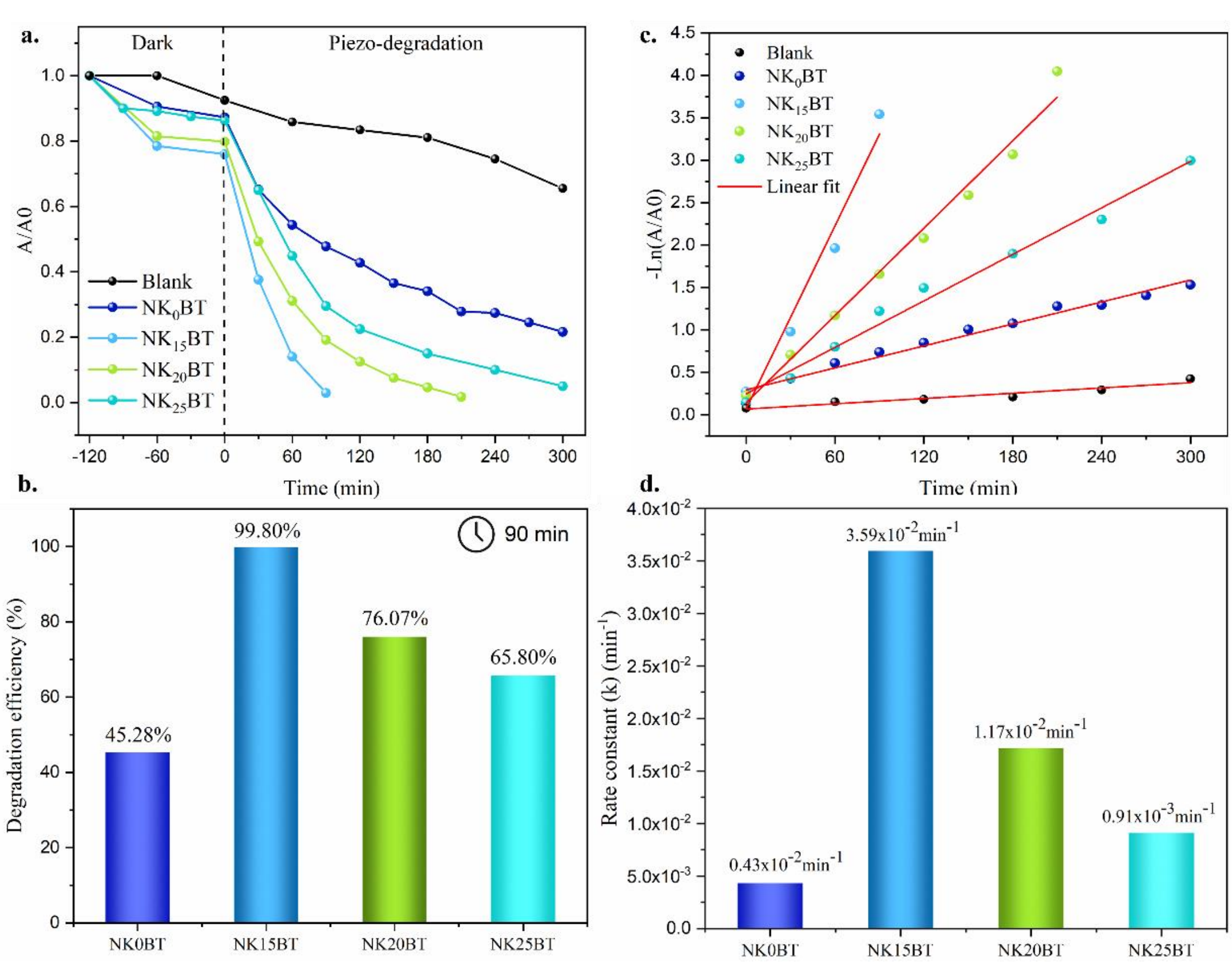

***Figure 13. a.*** *Plots of A/A0 versus time using NK$_x$BT piezocatalysts,* ***b.*** *Plots of -Ln(A/A0) versus time,* ***c.*** *Degradation efficiency,* ***d.*** *Rate constant of of NK$_x$BT piezocatalysts.*

***Table 2.*** *Comparison of piezocatalytic performance of various piezocatalysts.*

| Catalyst | Dye | [Dyes] (mg/L) | Catalyst dosage (g/L) | Ultrasonic source | Degradation efficiency | Rate constant k (min$^{-1}$) | Ref. |
|---|---|---|---|---|---|---|---|
| **NBT Micron crystals** | RhB | 10 | 4 | 40KHz/120W | 59%,180 min | 0.0044 | [46] |
| **NBT nano-particles** | RhB | 10 | 2 | 40KHz/150W | 92%,120min | 0.0220 | [52] |
| **$(Na_{0.8}K_{0.2})_{0.5}Bi_{0.5}TiO_3$** | MB | 10 | 2 | 40KHz/200W | 94%,160 min | 0.0169 | [45] |
| **$BaTiO_3$** | MO | 5 | 1 | 40KHz/80W | 98%,160 min | 0.0150 | [53] |
| **$0.3Ba_{0.7}Ca_{0.3}TiO_3$-$0.7BaSn_{0.12}Ti_{0.88}O_3$** | RhB | 6 | 1 | 37KHz/320W | 90%,120 min | 0.0195 | [54] |
| **NK$_{15}$BT** | RhB | 5 | 1 | 37KHz/300W | 98%,90min | 0.0359 | This work |

## 8.Understanding the mechanism of NK$_{15}$BT

To understand the active species involved in the degradation processes and contributing to the superior efficiency of NK$_{15}$BT across the piezo-catalytic system, scavenger tests were also conducted. The main motive of the scavenger study is to find the major reactive species during the evaluation of the piezo-degradation processes of RhB dye using the NK$_{15}$BT sample. The reactive species such as holes ($h^+$), superoxide anion radical ($\cdot O_2^-$), electrons, and hydroxyl free radical ($\cdot OH$) are trapped by using ethylenediaminetetraacetate (EDTA),benzoquinone (BQ), dimethyl sulfoxide (DMSO) and isopropyl alcohol (IPA) scavengers respectively [55]. Figure 14 shows the result regarding RhB degradation of $NK_{15}BT$ with and without scavengers.

With the addition of EDTA and IPA, it can be seen that a greater drop in the piezo-catalytic degradation efficiency occurred (Figure 14). Therefore, holes and hydroxyl radicals are considered as "effective primary" active species in the degradation mechanism, as estimated by CB and VB positions of $NK_{15}BT$. Furthermore, electrons and superoxide radicals have been shown to play a minor role in RhB decomposition.

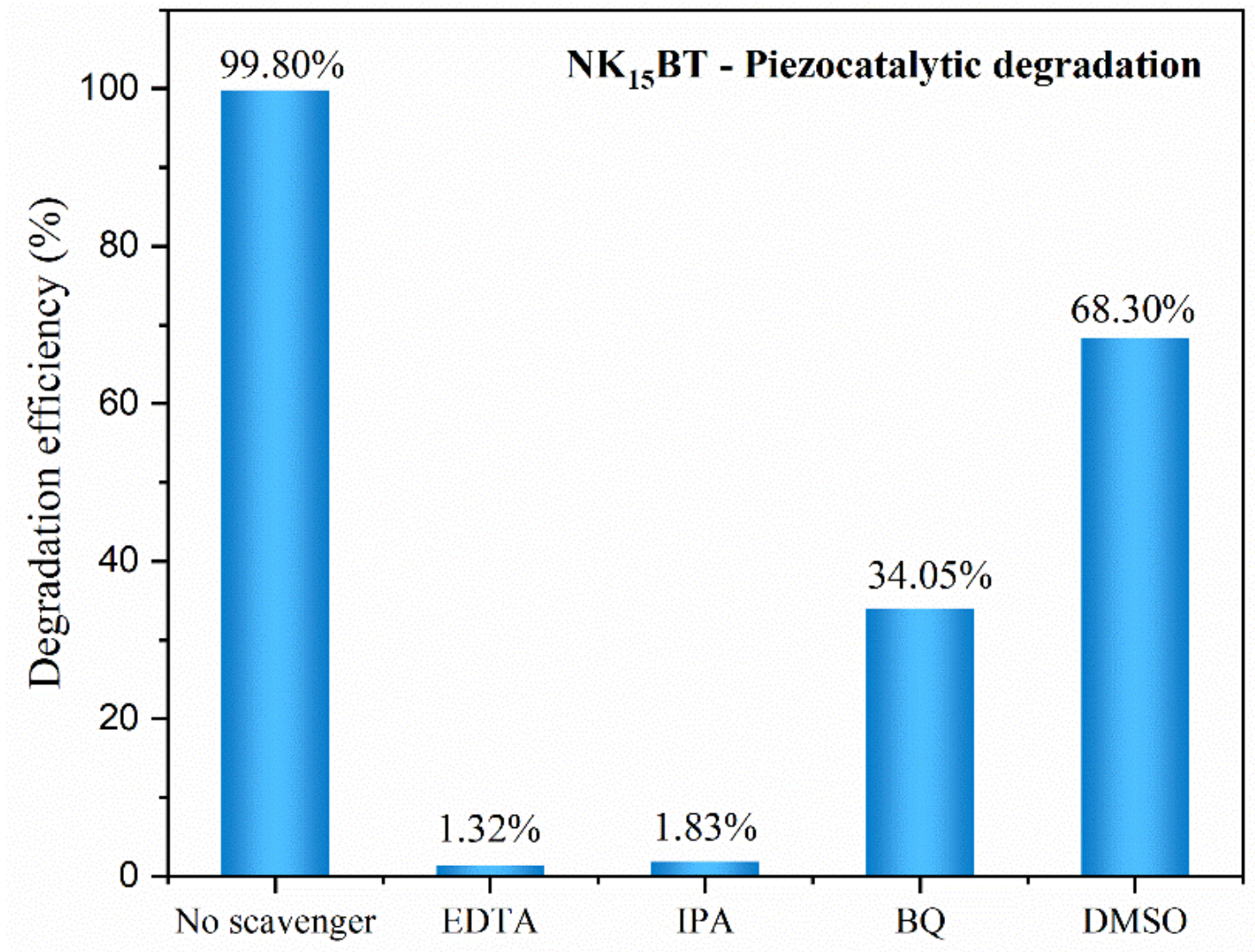

***Figure 14.*** *Radicals' scavenger test over $NK_{15}BT$ in the RhB piezo-degradation process.*

## 9. Recycling

The recyclability of a catalyst is of great importance from the practical perspective of wastewater treatment applications. Therefore, the recyclability of $NK_{15}BT$ catalyst was evaluated. After each cycle the catalyst was recovered by centrifugation and washed with distilled water and ethanol. As presented in Figure 15, $NK_{15}BT$ remains high degradation efficiency after 3 cycles. The piezo-degradation efficiency of RhB after the third cycle was found to be slightly reduced to 86.54%, which result from the loss of catalyst during washing and centrifugation resulted from the incomplete recovery of the initial mass of the catalyst.

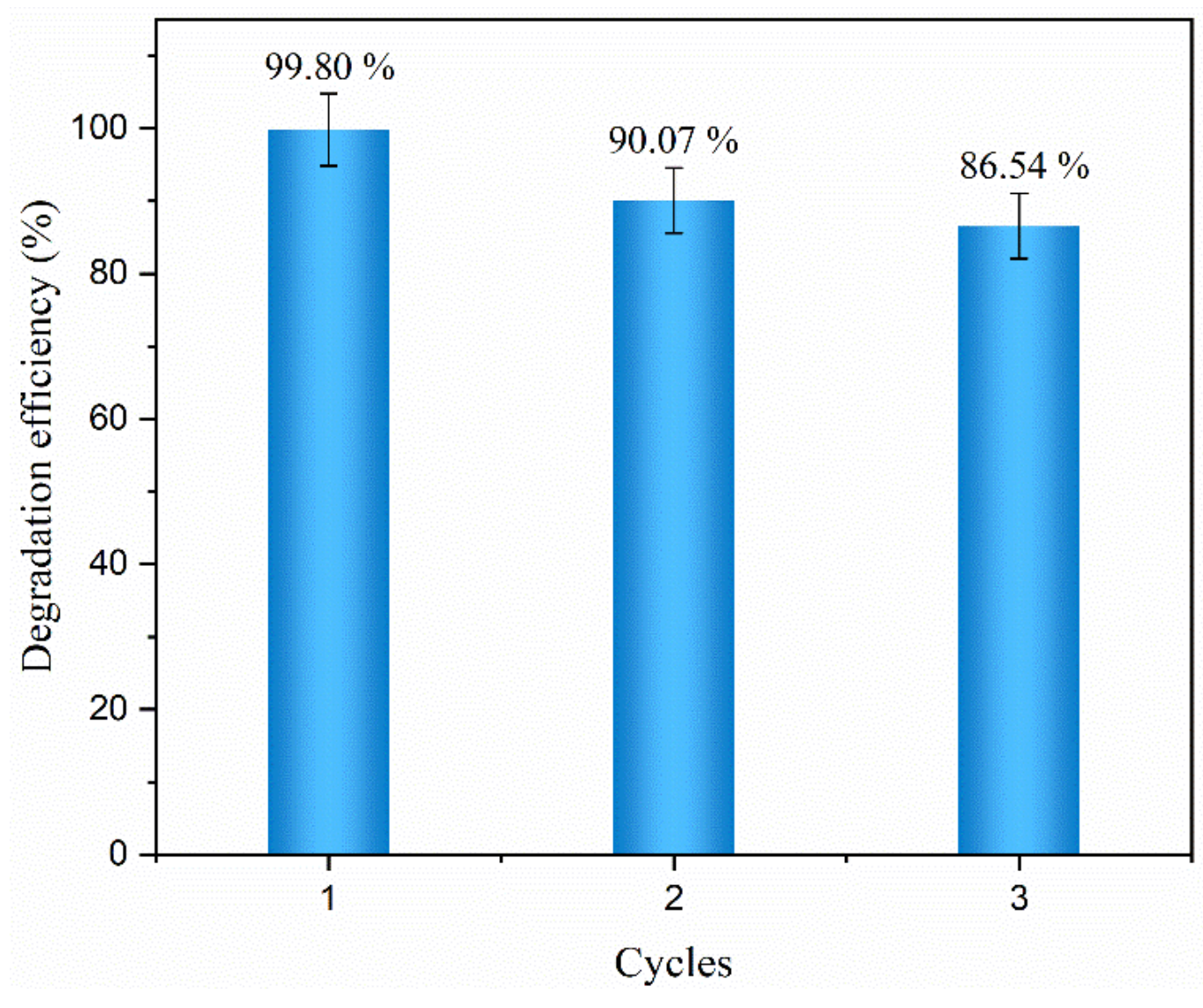

***Figure 15.*** *Recycling test of $NK_{15}BT$ catalyst.*

## 10. Total organic carbon (TOC) analysis

TOC analysis was performed to evaluate the state of conversion to $H_2O$ and $CO_2$ through the catalytic degradation process for all compositions in piezo-degradation. Figure 16 shows the TOC removal using the K-doped NBT catalysts in the piezo-degradation of RhB. The removal of TOC from the piezo-catalytic reaction reached 52.00%, 60.53%, 59.75% and 56.09% for $NK_0BT$, $NK_{15}BT$, $NK_{20}BT$ and $NK_{25}BT$, respectively. These results confirmed the excellent mineralization of RhB by the $NK_xBT$ piezocatalysts.

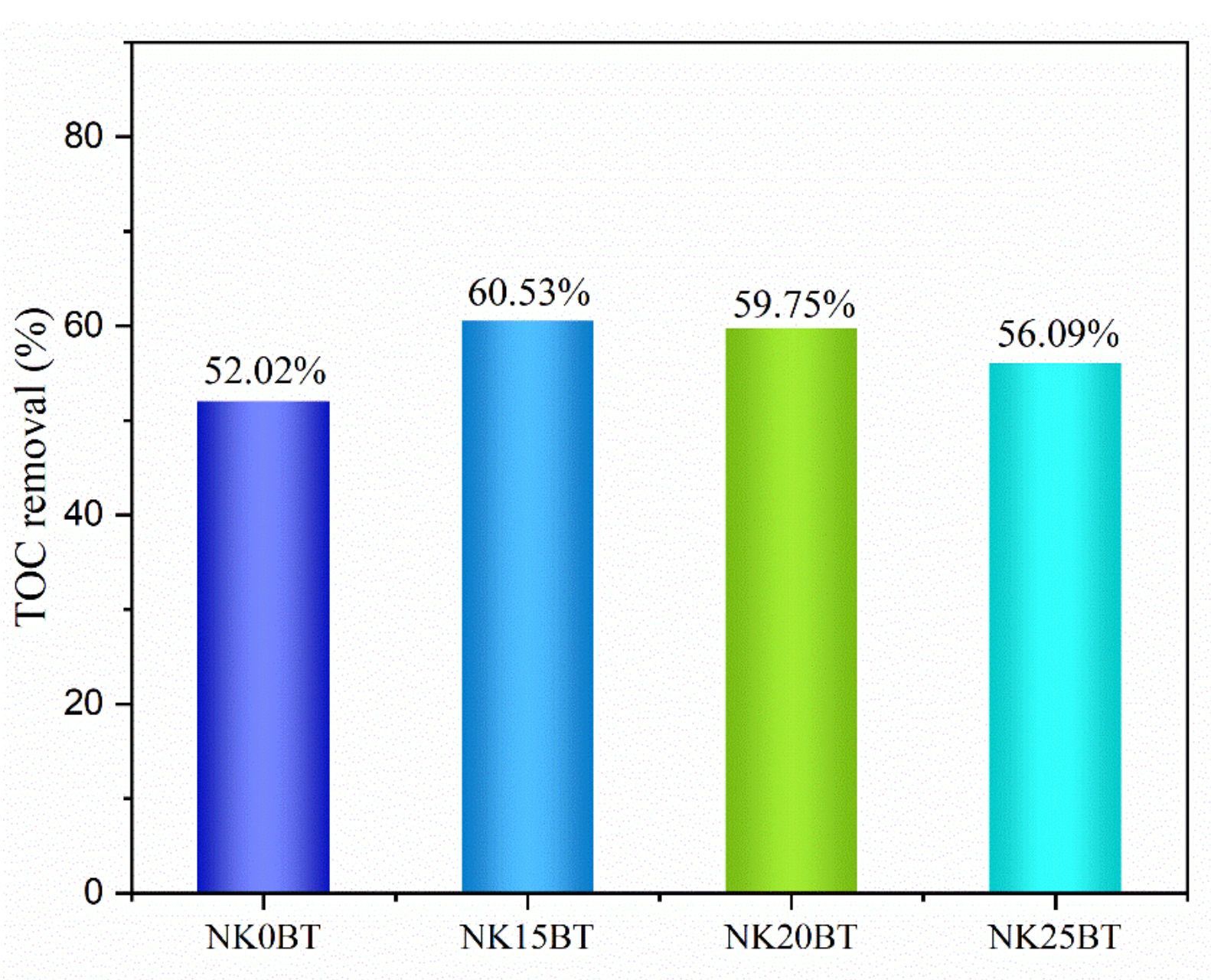

***Figure 16.*** *TOC removal by NKxBT piezo-catalysts.*

## Conclusion

In this work, a sol-gel auto-combustion method was used to synthesize the NKxBT catalysts with the composition x=0, 15, 20, 25. Subsequently, the structural properties of catalysts were characterized using XRD, HRTEM and Raman spectroscopy. XRD and HRTEM analyses revealed a perovskite structure dominated by the pseudo-cubic phase. Raman spectroscopy proved the coexistence of other phases, as suggested by XRD results; the tetragonal and pseudo-cubic phases in all compositions, and also their coexistence with the rhombohedral phase, observed only in $NK_{15}BT$. HRTEM images indicated a cuboid-like morphology and a decrease in grain size with $K^+$-doping as well. Optical properties investigated through bandgap evolution, revealed that the catalyst with the composition $NK_{15}BT$ had the lowest bandgap value. Thereafter, the piezo-catalytic activity of the synthesized samples was evaluated under ultrasonic vibration against RhB as model dye. The synthesized piezo-catalysts have shown noticeable performance, compared to pure NBT, yielding a maximum of dye degradation 99.8 % for $NK_{15}BT$. This improvement was correlated to its optical properties (optimal band gap) coupled with small particle size, and the co-existence of the three phases. Radical scavenging experiments validated that the piezo-generated holes and ·OH radicals were the two main active species which were responsible for piezo-catalytic degradation by $NK_{15}BT$. The $NK_{15}BT$ sample demonstrated good stability, as its degradation efficiency remained high after three consecutive recycling tests. Furthermore, TOC results revealed good mineralization of the RhB dye using $NK_xBT$ piezo-catalysts.